\documentclass[pra, 10pt, letterpaper,oneside, twocolumn]{revtex4-1}

\usepackage[latin1]{inputenc}
\usepackage{mathtools}
\usepackage{amsfonts}
\usepackage{array}
\usepackage[T1]{fontenc}
\usepackage{amssymb}
\usepackage[left=0.5in, right=0.5in, top=1in, bottom=1in]{geometry}

\usepackage{float}
\usepackage{graphicx}

\newcommand{\bibs}{C:/Users/seanm_000/Dropbox/References/BibFile}
\newcommand{\E}{\mathcal{E}}

\begin{document}
\author{Dennis Sullivan}
    \affiliation{Department of Electrical and Computer Engineering, University of Idaho, Moscow, Idaho 83844-1023}
    \email{dsulliva@uidaho.edu}
\author{Sean Mossman}
    \affiliation{Department of Physics and Astronomy, Washington State University, Pullman, Washington  99164-2814}
    \email{sean.mossman@wsu.edu}
\author{Mark G. Kuzyk}
    \affiliation{Department of Physics and Astronomy, Washington State University, Pullman, Washington  99164-2814}

\title{Hybrid quantum systems for enhanced nonlinear optical susceptibilities}


    \begin{abstract}
    Significant effort has been expended in the search for materials with ultra-fast nonlinear-optical susceptibilities, but most fall far below the fundamental limits.  This work applies a theoretical materials development program that has identified a promising new hybrid made of a nanorod and a molecule.  This system uses the electrostatic dipole moment of the molecule to break the symmetry of the metallic nanostructure that shifts the energy spectrum to make it optimal for a nonlinear-optical response near the fundamental limit.  The structural parameters are varied to determine the ideal configuration, providing guidelines for making the best structures.
    \end{abstract}

\maketitle

\section{Introduction}
    Nonlinear optical technologies allow for fine control over the phase, frequency and polarization of light that cannot be achieved through any other means\cite{boyd09.01}. Ultra-fast applications require the electronic mechanism, which originates in the virtual deformation of the electron cloud\cite{chang14.01}.  The magnitude of the nonlinear optical response of microscopic systems has grown steadily, but due to the suboptimal use of the electron oscillator strength in reported quantum system, the response is still orders of magnitude from the fundamental limits.\cite{kuzyk01.01, kuzyk13.01}.

    While many investigations into microscopic nonlinear polarizabilities have focused on molecular systems, and in larger plasmonic-type systems\cite{litch15.01} at the other extreme, combinations of such systems have not been systematically studied.  We propose that breakthroughs can be made by considering a class of hybrid materials that are based on coupling small molecular systems with nanostructures as they approach the quantum size regime.

    Monte Carlo investigations have shown that the maximum intrinsic nonlinear responses require energy differences which scale at least as $n^2$, termed superscaling\cite{shafe11.01}. Nanostructures are an example of a class of systems that exhibit superscaling characteristics. However, superscaling is not a sufficient condition.  In addition, the electronic transition strengths must also be tailored to optimize wave function overlap between states, requiring additional features to be added to the underlying electronic potential. Interactions with a dipolar molecular system as a source of an external electric field can provide nontrivial electron dynamics within the framework determined by nanostructure boundary conditions.

    This new paradigm for device fabrication could result in considerable impact on future development of nanostructure materials for nonlinear optical applications, as well as motivate efforts toward developing plasmonic devices approaching the quantum regime.

    Our theoretical materials discovery program has found that the hyperpolarizability of a cylindrical nanowire with an aspect ratio of at least 10:1, which is attached to a permanent electric dipole with an inert spacer bridge, yields a first hyperpolarizability near the fundamental limit.  The ideal configuration is found by varying the dipole strength and distance between the molecule and nanowire while calculating the off-resonant electronic response of an electron using the finite-difference time-domain (FDTD) technique\cite{sulli16.01}.  While many theoretical models yield a hyperpolarizability near the limit, the system proposed here is within reach of modern nano-fabrication techniques.

    \begin{figure}
        \includegraphics[width=\linewidth]{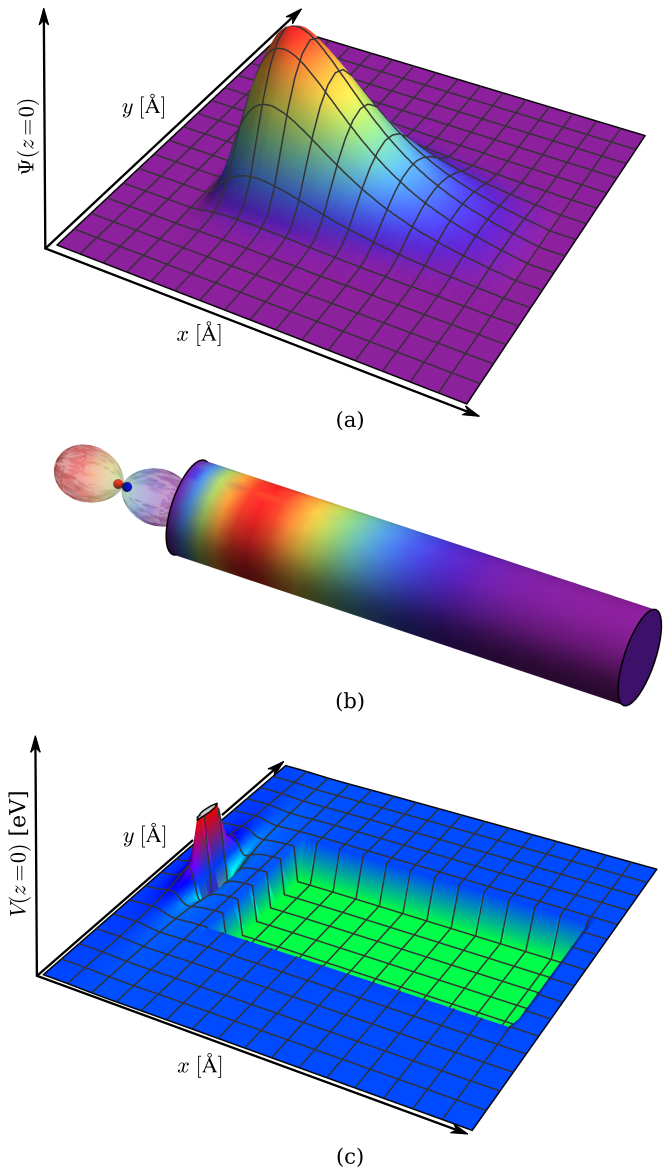}
        \caption{The proposed nanostructure is displayed from three perspectives: (a) a two dimensional cross-section of the single particle wavefunction; (b) a three dimensional representation of the physical system with a wave function contour; and (c) a two dimensional cross-section of the underlying electronic potential used to model the molecule-nanowire structure. }
        \label{fig:colfig}
    \end{figure}

\subsection{Hyperpolarizabilities}

    When the applied electric field is small, the polarization response to an optical field can be approximated by the expansion
    \begin{equation}
        P_i = P_i^{(0)}+\chi^{(1)}_{ij}\E_j+\chi^{(2)}_{ijk}\E_j\E_k+\dots
        \label{eq:polarization}
    \end{equation}
    where $i,j,k$ are Cartesian tensor components of the external electric field $\E_i$ and the resulting polarization of the material, $P_i$. The nonlinear susceptibilities $\chi^{(\mu)}$ completely characterize a material's response to an external field in the dipole approximation and will depend on the arrangement, number density, and fundamental response of the microscopic systems which make up the material -- to name a few.

    Nonlinear optical technologies which respond on the fastest timescales rely on the electronic response of the microscopic constituents, which form the bulk materials. These microscopic responses are defined by an expansion similar to Eq. \ref{eq:polarization} but for the quantum expectation value of the molecular dipole moment,
    \begin{equation}
        \mu(\E) = e\langle x\rangle = \mu_0 + \alpha_{ij}\E_j+\beta_{ijk}\E_j\E_k+\dots,
        \label{eq:dipexp}
    \end{equation}
    where $\alpha$ is the polarizability and $\beta$ is the first hyperpolarizability.  When these units are on molecular size scales, the nonlinear coefficients are calculated from the quantum expectation value of the molecular dipole moment in the presence of the electric filed.  As in the macroscopic case, the expansion coefficients fully characterize the microscopic units.  Maximizing the first hyperpolarizability is the goal of this work.

\subsection{Characteristics of the optimum hyperpolarizability}

    The quantum mechanical oscillator strength is the fundamental measure of the strength of light-matter interactions and is limited by the Thomas-Reich-Kuhn sum rules\cite{bethe77.01}, which can be derived directly by evaluating the commutator $ \langle l|[x,[x,H]]|p \rangle$, which yields
    \begin{equation}\label{eq:sumrules}
        \sum_{n} x_{ln}x_{np}\left(E_n-\frac{1}{2}(E_l+E_p)\right) = \frac {\hbar^2N} {2m}\delta_{lp} ,
    \end{equation}
    where $N$ is the number of electrons, $m$ their mass, and the mechanical Hamiltonian is of the form $H = p^2/2m + V(\mathbf{r})$.  Note that Eq. \ref{eq:sumrules} holds for any quantum system described by a mechanical Hamiltonian.

    It has been shown\cite{kuzyk00.01} that these sum rules can be used to determine a limit on the hyperpolarizabilities under the assumption that the nonlinear response is maximized when the oscillator strength is concentrated amongst three states.  This assumption along with Eq. \ref{eq:sumrules} determine a maximum hyperpolarizability allowed by quantum mechanics to be
    \begin{equation}
        \beta_\text{max} = \sqrt[4]{3}\left(\frac{e\hbar}{\sqrt{m}}\right)^3\frac{N^{3/2}} {E_{10}^{7/2}}
        \label{eq:betamax} ,
    \end{equation}
    where $E_{10}$ is the first energy difference $E_1-E_0$. For the remainder of this work all calculations of the hyperpolarizability will be normalized to this maximum and therefore represent the intrinsic value $\beta_\text{int} = \beta/\beta_\text{max}$, which is invariant under a global length scale change.

    Through extensive potential optimization\cite{zhou06.01,burke13.01,lytel13.01} it has become clear that there exists an apparent limit to the hyperpolarizability of real systems which is $0.7089\beta_\text{max}$, while molecules engineered for nonlinear-optical applications are often a factor of 30 below the fundamental limit. However, by sampling random transition moments and energy spectra constrained only by the sum-rules, Shafei, et al.,\cite{shafe11.01} showed that the fundamental limit is achievable in principal only by energy spectra which scale as $n^2$ or faster. Additionally, a large hyperpolarizability requires a careful balance of charge transfer and wave function overlap in the first few populated states. Applying an external field to the system is one way to tune these characteristics of the system. Thus, a quantum confined system provides the necessarily This motivates the use of a nanowire to attain such a spectrum.  The static dipole, on the other hand, affects the overlap between the wave functions without a significant effect on the energy spectrum.

\section{Hybrid Nanowire and Dipolar Molecule}

    The choice of hybrid system is motivated by the requirements of a large nonlinear-optical response -- attempting to take advantage of the superscaling behavior of quantum confined systems, the charge transfer of long and narrow structures, and the tantalizing scaling characteristics of the $1/x^2$ potential. The desirable $n^2$ energy spectrum is achieved in a cylindrical nanowire by confining electrons on a quantum scale. A permanent electric dipole produces an effective $1/x^2$ potential and provides additional controllable parameters that break the symmetry of the electron wavefunctions and allow for optimized transition characteristics.

    This system can be produced in practice using a cylindrical metallic nanowire, which we fix at 10\AA\ in diameter and 100\AA\ in length so that the dominant excitations are concentrated along the applied electric field; and, a molecule with a permanent electric dipole in close proximity to one end as schematically shown in Fig. \ref{fig:colfig}. The distance and strength of the physical dipole, modelled as two point charges of opposite charge, is varied to determine where the response of the system is optimized.

\begin{figure}
        \includegraphics[width=\linewidth]{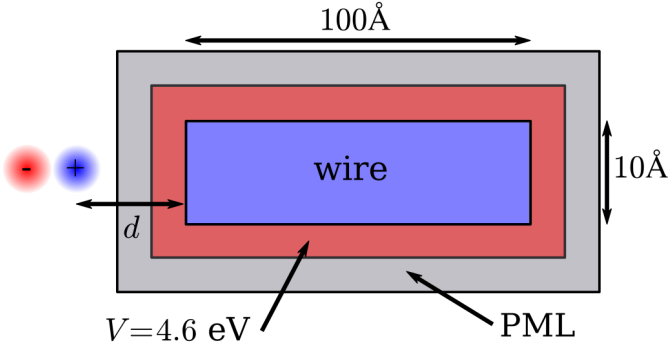}
        \caption{A finite cylindrical well in a 120x30x30 problem space with well dimensions of 100x10x10 cells, where each cell is $1 \mbox{\AA}^3$ volume. A perfectly matched layer prevents probability amplitude from reflecting back into the problem space. The dipole need not be included in the problem space as its static field affects the electron in the well without contributing to the dynamics.}
        \label{fig:simdiagram}
    \end{figure}

    We simulate this structure in full 3D as a single electron in an finite cylindrical well, taking the depth to be the work function of silver, though the result is rather insensitive to the precise depth of the well. The ground state solution, physical system, and spatial potential are shown in Fig. \ref{fig:colfig} and the problem space is shown schematically in Fig. \ref{fig:simdiagram}. The outermost boundary of the problem space consists of a 3 \AA\ thick perfectly matched layer (PML) as described in Appendix \ref{sec:FDTD}. A cross section of the ground state wavefunction is shown in Fig. \ref{fig:colfig}a.

\section{Results and Discussions}

    The Schrodinger Equation is solved using Finite Difference Time Domain (FDTD), a technique described in Appendix \ref{sec:FDTD}.  Appendix \ref{sec:CHO} applies the technique to the clipped harmonic oscillator, which has analytical solutions.  The numerical results produced using FDTD agree with the analytical solutions, showing that the technique's viability in calculating nonlinear susceptibilities. Appendix C\ref{sec:FF} describes how the hyperpolarizability is calculated from the expectation value of the dipole operator using the finite field method.

    These techniques were applied to the proposed hybrid system.  The ground state and first excited state eigenvalues in the absence of an applied field are found to be $E_0 = 0.5896$ eV and $E_1 = 0.6053$ for a 9.6 D dipole at a separation distance of 16 \AA.  Next, the induced dipole moment as a function of the applied electric field is determined, from which the hyperpolarizability is calculated by taking the second derivative.  The optimum static field range for getting accurate derivatives was determined to be from -0.025 to 0.025 mV/\AA, within which 5,000 dipole moment data points were taken and the hyperpolarizability calculated.

    \begin{figure}
    \includegraphics{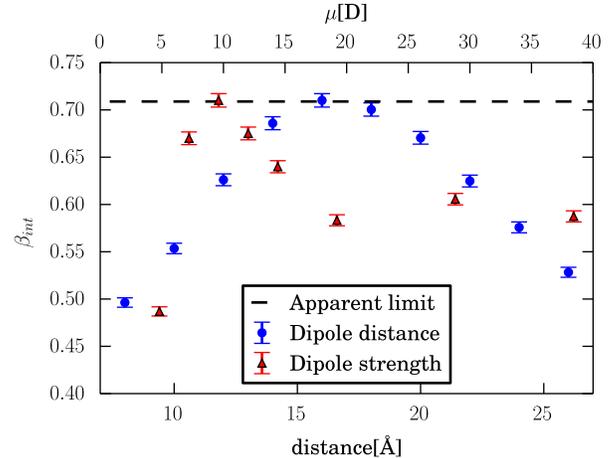}
    \caption{The intrinsic hyperpolarizability as a function of the dipole strength (when the distance is 16 \AA) and the distance of the dipole from the end of the wire (when the dipole moment is 9.6 D) showing that the system can be optimized to yield the apparent limit.}
    \label{fig:betawiredipole}
    \end{figure}

    This calculation was repeated for a range of dipole moment strengths, which were varied by adjusting the point charges representing the molecule, and the distance between the dipoles and the edge of the nanowire.  The data generated are shown in Fig. \ref{fig:betawiredipole}.  The hyperpolarizability peaks when $\mu = 9.6 D$ and $d = 15 \mbox{\AA}$.

    There are two significant features of note; the hyperpolarizability peaks and the peak value is at the apparent limit.  All measured molecules and quantum systems -- with the exception of small twisted molecules reported by Kang et al\cite{kang05.01,Kang07.01} -- fall far short of this upper bound; and, it is not clear if the hyperpolarizability scales favorable in the best existing molecules.  Our results suggest that any molecule would do in combination with the nanowire provided that the primary excitation is in the nanowire and that the dipole moment and its distance from the wire meets the conditions at the peaks shown in Fig. \ref{fig:betawiredipole}, i.e. where $\mu = 9.6 D$ and $d = 15 \mbox{\AA}$.  As such, this hybrid system is a highly flexible paradigm for the fundamental unit of nonlinear response that should scale favorably.

\begin{figure}
    \includegraphics{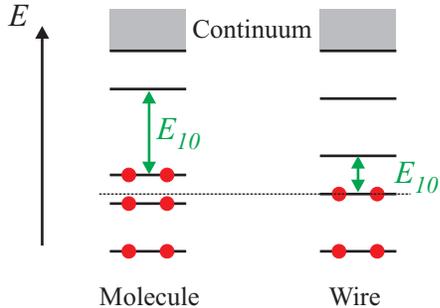}
    \caption{The energy difference between ground and first excited state of the molecule must be greater that of the wire and there should be no available states for charge transfer between the two.}
    \label{fig:Energy}
    \end{figure}

    It is important to stress the conditions of validity of our model.  First, the dipole moment must be static and the primary excitation must reside in the nanowire.  For this to be the case, the energy difference between the first excited state and ground state of the molecule must be larger than that for the nanowire.  Fig. \ref{fig:Energy} shows an example of an energy-level diagram that meets the requirements.

    Implicit in these discussions is the requirement that the wire be small enough to support quantum excitations.  Also, the energy levels must be arranged in a way that there cannot be charge transfer between the two.  For example, in Fig. \ref{fig:Energy}, the highest occupied level in the molecule is higher than in the wire, but because its state is filled with the maximum number of electrons, Pauli exclusion prevents charge transfer.

    Finally, the most blaring deficit of our calculations may appear to be in the fact that only one electron is included.  Clearly, one cannot accurately calculate the nonlinear response of a real system with one electron.  However, it has been shown that the peak values of the nonlinear response are unchanged when another electron is added, nor when interactions between them are included.\cite{watki11.01} Therefore, the important result is that a peak exists, though it will most likely not be at the predicted values given here.  Thus, our work is intended to show that a fabrication effort of such hybrid systems based on dipoles and nanwires is likely to be fruitful in producing quantum systems that have record nonlinear response.

\section{Real Systems}

    There are many possible ways to implement the hybrid systems that we propose here, though some methods may require a more sophisticated analytical approach to model.  For example, density functional theory could be used to treat systems with fabrication methods that rely on the adsorption of a molecule onto the surface by van der Waals forces or ionic binding. The latter would employ charge transfer that would drastically effect both the Fermi level and the static dipole moment.

    We suggest the use of less invasive methods such as employing an inert spacer group -- which is electronically stable, allows for binding to metals such as silver without complication, and trivializes the problem of relative Fermi level imbalance.  For example, it has been known for over 6000 years that silver tarnishes, a chemical process that leads to a coating of silver.  Silver's affinity for sulfur can be used as a method for anchoring a spacer group to the end of a nanowire.

    Fu and Lakowicz, for example, demonstrated that fluorescing molecules could be attached to 50nm diameter spherical silver nanoparticles using sulphur at the attachment point.\cite{fu07.03}  The fluorescence yield of an attached fluorophore was shown to be enhanced 15-fold over the single molecule.  We stress that our proposal for the nanowire hybrid differs greatly from the work of Fu.  In our case, the excitations take place in the metal, the response is quantum in nature, and the role of the molecule is passive.  Fu's system, on the other hand, uses the classical surface plasmon to enhance the light in the vicinity of the molecule.  It has been demonstrated that classical effects take hold for particle sizes above 20nm,\cite{scholl12.01} so our system -- being a factor of two below the threshold size -- should exhibit the required quantum behavior.

    The magnitude of the required dipole moment is within the range of typical organic molecules, and tethers with sulphur end groups are commonly synthesized, so there should be no obstacles to making nanowire-molecular composites whose nonlinear-optical response attains the apparent limit.

\section{Conclusions}

    We have demonstrated through FDTD computations that a hybrid quantum system made of a 100 \AA\ silver nanowire and connected with a tether to a molecule with suitable dipole moment can lead to a hyperpolarizability at the apparent limit.  The prospect for such a large nonlinearity makes fabrication investigations worthwhile to identify the ideal structural parameters for attaining this limit.

    \section{Acknowledgements}
        SM and MGK thank the National Science Foundation (ECCS-1128076) for generously supporting this work.

\section{Appendix}

\subsection{A. Finite Difference Time Domain}\label{sec:FDTD}

    To  determine the electronic ground state for a particular system, we employ a finite difference method which uses the time evolution of an initial pulse to determine the stationary state solutions.\cite{sulli16.01} The FDTD method benefits from being computationally efficient while being fully generalizable for solving the Schr\"odinger equation for arbitrary potentials in three dimensions.

    We begin with the time-dependent Schr\"odinger equation
    \begin{align}
        -i\hbar\frac{\partial\psi(x,y,z;t)}{\partial t} = \frac{\hbar^2}{2m}&\nabla^2\psi(x,y,z;t)\nonumber\\
        &+V(x,y,z;t).
    \end{align}
    Separating the solution into real and imaginary parts produces two coupled equations:
    \begin{subequations}
    \begin{align}
        \frac{\partial\psi_\text{real}(x,y,z,t)}{\partial t} = -\frac{\hbar}{2m_e} &\nabla^2\psi_\text{imag}(x,y,z;t)\\
        &+\frac{1}{\hbar}V(x,y,z;t)\psi_\text{imag}(x,y,z;t),\nonumber\\
        \frac{\partial\psi_\text{imag}(x,y,z;t)}{\partial t} = \frac{\hbar}{2m_e} &\nabla^2\psi_\text{real}(x,y,z;t)\\
        &-\frac{1}{\hbar}V(x,y,z;t)\psi_\text{real}(x,y,z;t).\nonumber
    \end{align}
    \label{eqs:spaceandtime}
    \end{subequations}
    These equations are then discretized in space and time by choosing a finite grid of size $\Delta x$ and $\Delta t$. Once a cubic grid is defined by the choice of $\Delta x$, a $\Delta t$ is chosen to be small enough to maintain stability. Then alternating iterations of Eqns. \ref{eqs:spaceandtime} simulate the motion of the waveform in time\cite{sulli02.01, soria04.01, ren04.01,ren08.01,sulli05.01,sulli12.01}.

    It is also necessary to prevent the outgoing waveforms from being reflected by the boundaries back into the problem space without making the problem space unrealistically large. This is accomplished with a perfectly matched layer (PML)\cite{beren94.01,sulli12.02,zheng07.01}: a region within the simulation space which behaves like a perfect absorber, damping outgoing waveforms in such a way that does not affect the interior of the problem space. Under these boundary conditions, we are able to verify that our solutions are bound without interacting with the PML by monitoring the normalization of the solution over time iterations.

    The simulation is initialized by beginning with an arbitrary pulse and evolving it in time. This pulse is chosen arbitrarily except that it is not orthogonal to the true ground state solution that we are seeking. As time evolves, the pulse evolves and disperses as determined by its spectral decomposition in the eigenstate basis determined by the problem space potential. Some of the pulse will propagate into the PML, effectively leaving the problem space. The waveform which remains can be described as
    \begin{equation}
        \psi(x,y,z;t) = \sum_{n=0}^N\phi_n(x,y,z)e^{-iE_nt/\hbar}
    \end{equation}
    where $\phi_n(x,y,z)$ are the eigenfunctions and $E_n$ are the corresponding eigenenergies. The eigenenergies can be determined by monitoring the time-domain data at a single point, $r_0$, and taking the Fourier transform
    \begin{align}
        \mathfrak{F}\left\{\psi(r_0;t)\right\}&=\int_{-\infty}^\infty dt \left[\sum_{n=0}^N\phi_n(r_0)e^{-iE_nt/\hbar}\right]e^{i\omega t}\\
        &=\sum_{n=0}^N\phi_n(r_0)\delta\left(\omega-E_n/\hbar\right)
    \end{align}
    producing a series of delta functions in the frequency domain corresponding to the eigenenergies of the system in question. The eigenfunctions are then recovered by taking a discrete Fourier transform waveform at the frequency $\omega_m=E_m/\hbar$ at every point in the problem space:
    \begin{align}
        \text{DFT}&\left\{\psi(r;t)\right\}_{\omega_m}=\int_{-\infty}^\infty dt \left[\sum_{n=0}^N\phi_n(r)e^{-iE_nt/\hbar}\right]e^{iE_mt/\hbar}\\
        &=\int_{-\infty}^\infty dt \left[\sum_{n=0}^N\phi_n(r)e^{-i(E_n-E_m)t/\hbar}\right]=\phi_m(r).
    \end{align}

    \subsection{B. Calculating the hyperpolarizability in a clipped harmonic oscillator} \label{sec:CHO}

    In this section, we apply the FDTD method described above to the clipped harmonic oscillator and use those solutions to calculate the first hyperpolarizability $\beta$. A clipped harmonic oscillator is a three-dimensional harmonic oscillator centered at zero but confined to the positive $X$, $Y$, and $Z$ octant. We use this potential because an analytic solution is known with which we may determine the accuracy of the method.

    \begin{figure}[H]
        \centering
        \includegraphics[width=0.7\linewidth]{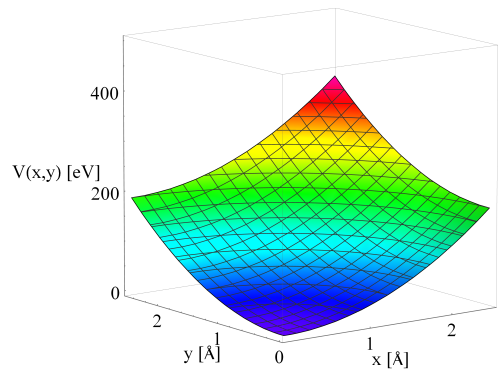}
        \caption{The potential in the $X$ and $Y$ directions of a three-dimensional clipped harmonic oscillator defined by a 30 eV energy scale. The planes at $x=0$, $y=0$, and $z=0$ are fixed at infinite potential. The boundaries at $x$, $y$, and $z=2.5$\AA\ have a three cell PML.}
        \label{fig:CHOpotential}
    \end{figure}

    We limit ourselves to a diagonal tensor component of the hyperpolarizability, as will be of most interest for the wire system, by taking all polarizations and fields to be along one cartesian direction. We are looking to calculate
    \begin{equation}
        \beta = \left.\frac{1}{2}\frac{\partial^2 p(\mathcal{E})}{\partial\mathcal{E}^2}\right|_{\mathcal{E}\to 0}
        \label{eq:beta}
    \end{equation}
    where $\mathcal{E}$ is a constant electric field across the problem space. We calculate the above derivative directly by computing the ground state dipole moment for a variety of electric field strengths, extract the second order fitting parameter, and express the result in intrinsic units as mentioned following Eq. \ref{eq:betamax}. The maximum hyperpolarizability can be expressed in these units as
    \begin{equation}
        \beta_\text{max} = 27.68278(\epsilon_1-\epsilon_0)^{-7/2} \left[\frac{e\text{\AA}^3}{\text{V}^2}\right]
        \label{eq:betamax_units}
    \end{equation}
    where $\epsilon_0$ and $\epsilon_1$ are the ground state energy and first excited state energy, respectively, in eV.

    \begin{figure}
        \includegraphics[width=\linewidth]{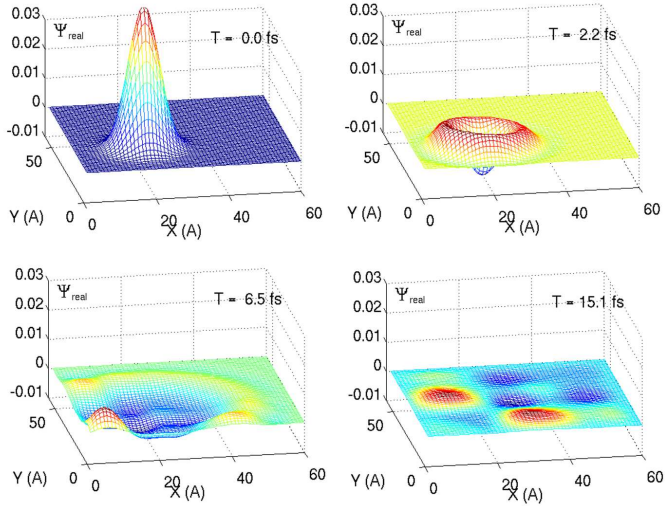}
        \caption{A test function is initialized in the problem space for the clipped harmonic oscillator. As the FDTD method proceeds, the waveform spreads out.}
        \label{fig:clippedHOinit}
    \end{figure}

\begin{figure}
        \includegraphics[width=\linewidth]{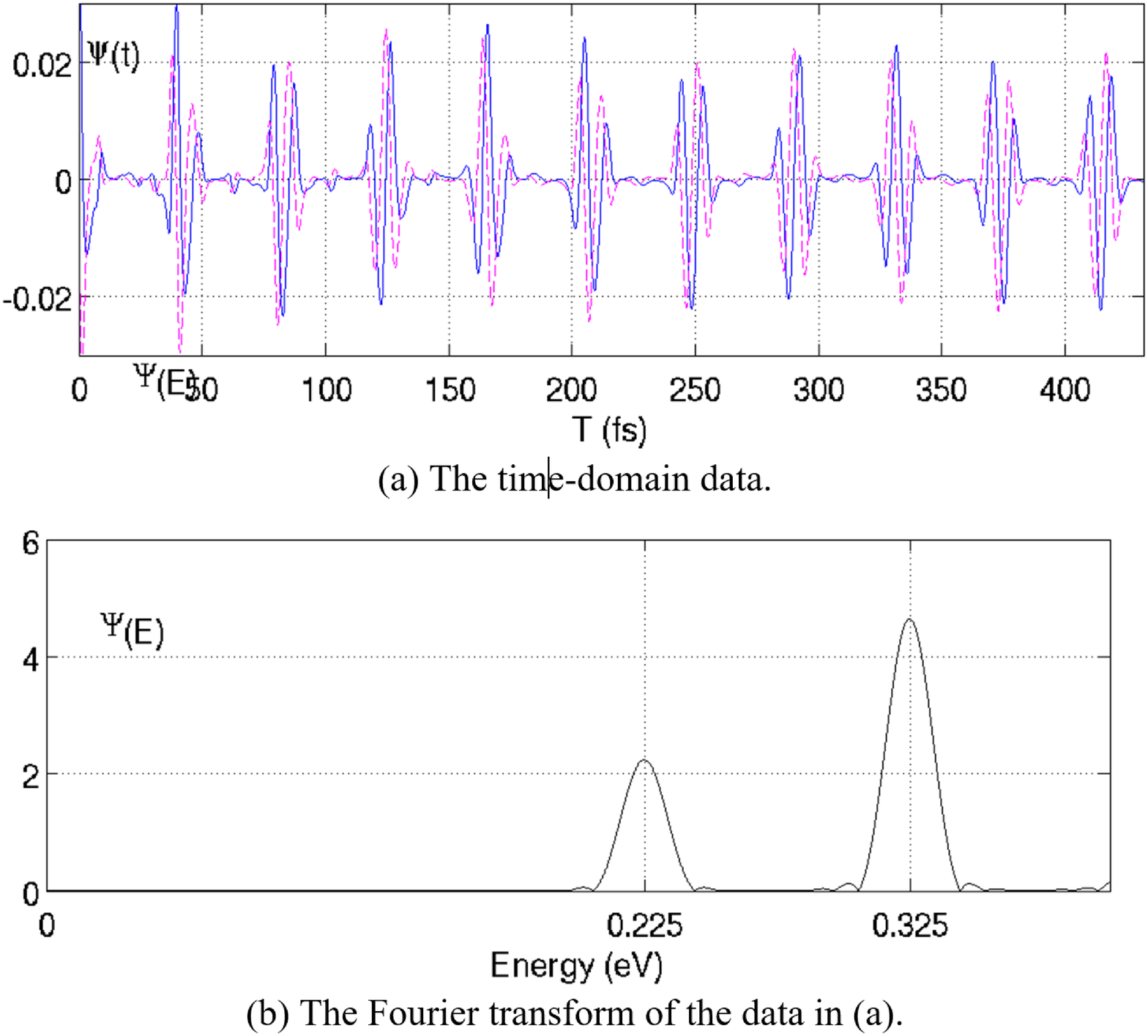}
        \caption{(a) The stored time-domain data for the real and imaginary amplitude at the initialization point. (b) The Fourier transform. The energies are negative.}
        \label{fig:clippedHOtimedata}
    \end{figure}

    The first step is to find the energies $\epsilon_0$ and $\epsilon_1$ that will be needed in Eq. \ref{eq:betamax_units}. We start by initializing a test function as shown in Fig. \ref{fig:clippedHOinit}. The simulation is run for 10,000 iterations, storing the time-domain data at the point where the test function was initialized. The time-domain data and the corresponding Fourier transform are shown in Fig. \ref{fig:clippedHOtimedata}. The resulting eigenenergies are $\epsilon_0=0.225$ eV and $\epsilon_1=0.325$ eV.

\begin{figure}
        \includegraphics[width=\linewidth]{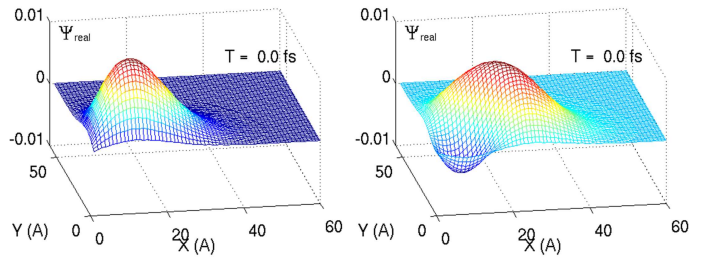}
        \caption{The ground and first excited wavefunctions for the clipped harmonic oscillator, corresponding to the energy peaks appearing in Fig. \ref{fig:clippedHOtimedata}b. }
        \label{fig:CHOstates}
    \end{figure}

    To construct the wavefunctions corresponding to the energy eigenvalues indicated by peaks in the Fourier transform, the original test function is initialized in the problem space as before, but as the simulation proceeds, a discrete Fourier transform at the frequency of the desired state is taken at every cell in the problem space. The process is repeated for each eigenstate we have identified. The results are shown in Fig. \ref{fig:CHOstates}.

    \subsection{C. Calculating the hyperpolarizability}\label{sec:FF}

        Now that we have established the FDTD method for determining the ground state wavefunction, we apply a finite fields (FF) algorithm to determine the first hyperpolarizability. To do this, we numerically determine the second derivative of the polarizability given in Eq. \ref{eq:beta} by evaluating the ground state wavefunction after applying a range of static electric fields to the problem space.

        We begin this process by applying a voltage of -0.5 mV/\AA\ along the long axis of the wire structure. The FDTD algorithm is initialized with the same test function as used previously and the ground state wavefunction is determined under the influence of the small static field. We then initialize the next simulation using the ground state wave function determined for the first static field measurement, then over 50,000 iterations the static field is varied from $-0.5$ to 0.5 mV/\AA. Every one hundred time steps the dipole moment is calculated to determine the curve $\mu(\E)$.

        To determine the most accurate value of the hyperpolarizability, we reduce the maximum magnitude of the electric fields applied to the system until numerical instabilities become apparent. For this harmonic oscillator, the optimum range of applied fields was found to be from -1 to 1 mV/\AA. The resulting curve is then fit to a fourth order polynomial, where the second order coefficient determines the first hyperpolarizability.

        Table \ref{tab:betaFF} shows the results of the analysis on the clipped harmonic oscillator in 3D with analysis to determine the effect of the numerical cell size. We determine the this method can determine the first hyperpolarizability to within 1\%.

        \begin{table}
            \begin{tabular}{|c|c|c|}
              \hline
              Cell Size [\AA]& $E_{10}=E_1-E_0$ [eV] & $\beta_\text{int}$ \\
              \hline
              1.0 & 0.09981 & 0.5716 \\
              0.9 & 0.09983 & 0.5717 \\
              0.8 & 0.09982 & 0.5716 \\
              0.7 & 0.09985 & 0.5720 \\
              0.6 & 0.09982 & 0.5714 \\
               -  & 0.1     & 0.5707 \\
              \hline
            \end{tabular}
            \caption{The finite field results for the 3D, clipped harmonic oscillator using the FDTD method of determining the ground state wave function. The sensitivity of the result to changing cell size is explored and compared with the fully analytic result.}
            \label{tab:betaFF}
        \end{table}


    \bibliography{\bibs}

\end{document}